\documentclass[12pt]{article}

\usepackage{etoolbox}

\newtoggle{ieeeversion}
\togglefalse{ieeeversion}

\usepackage[utf8]{inputenc} 

\usepackage[a4paper,top=2cm,bottom=2cm, left=2cm, right=2cm]{geometry}

\usepackage{graphicx} 

\iftoggle{ieeeversion}{%
}{%
\usepackage{booktabs} 
\usepackage{array} 
\usepackage{paralist} 
\usepackage{verbatim} 
\usepackage{subfig} 
}

\usepackage{nth}

\usepackage{ifpdf}
\usepackage{url}
\usepackage{epsfig,epic,color}
\usepackage{amsmath,amscd,amssymb,amsfonts,amsthm}
\usepackage{fancybox}
\usepackage{calc}
\usepackage{latexsym}
\usepackage{array}
\usepackage{ifthen}
\usepackage{multirow}
\usepackage{color}
\usepackage{colortbl}

\newtheorem{theo}{Theorem}[section]
\newtheorem{defi}{Definition}[section]
\newtheorem{coro}{Corollary}[section]
\newcommand{\eps}{\epsilon}
\newcommand{\gdo}[1]{\ensuremath{\mathcal{O}\left(#1\right)}}
\newcommand{\N}{\mathbb{N}}
\newcommand{\on}[1]{\operatorname{#1}}

\iftoggle{ieeeversion}
{%
}{%
\ifpdf
\fi

\usepackage{fancyhdr} 
\pagestyle{fancy} 
\lhead{}\chead{}\rhead{}
\lfoot{}\cfoot{\thepage}\rfoot{}

\usepackage{sectsty}
\allsectionsfont{\sffamily\mdseries\upshape} 

\usepackage{authblk}

}

\usepackage{hyperref}

\ifpdf
\hypersetup{
pdfauthor={Sébastien Kunz-Jacques}{Paul Jouguet},
pdftitle={Using Hash-Based Signatures to Bootstrap Quantum Key Distribution},
pdfsubject={cryptology, Quantum Key Distibution},
pdfkeywords={crytography} {quantum key distribution} {digital signatures} {hash functions} {post-quantum crypto},
pdfcreator={LaTeX},
pdfproducer={pdfLaTeX}
}
\fi

\title{Using Hash-Based Signatures to Bootstrap Quantum Key Distribution}

\iftoggle{ieeeversion}
{%
\author{S\'ebastien~Kunz-Jacques
        and~Paul~Jouguet%
\thanks{S. Kunz-Jacques is with SeQureNet, 23 avenue d'Italie, 75013 Paris, France (e-mail: sebastien.kunz-jacques@sequrenet.fr).}
\thanks{P. Jouguet is with
Institut Telecom / Telecom ParisTech, CNRS LTCI, 46, rue Barrault, 75634 Paris Cedex 13, France
and SeQureNet, 23 avenue d'Italie, 75013 Paris, France (e-mail: paul.jouguet@sequrenet.fr).}
}
}{%
\author[1]{Sébastien Kunz-Jacques}
\author[1,2]{Paul Jouguet}

\affil[1]{SeQureNet, 23, avenue d'Italie, 75013 Paris}
\affil[2]{Telecom ParisTech, 46, rue Barrault, 75013 Paris}

\date{} 
}
\begin{document}
\maketitle

\begin{abstract}
Quantum Key Distribution is a secret distribution technique that requires an authenticated channel. This channel is usually created on top of an un-authenticated communication medium using unconditionally secure Message Authentication Codes (MAC) and an initial common secret. We examine the consequences of replacing this MAC algorithm by a cryptographic hash-based signature algorithm, like the Lamport algorithm. We show that provided one-way functions exist, the Lamport algorithm or its variants can be instantiated in a secure way in the Universally Composable sense, and can therefore be plugged into any QKD protocol with a composable security proof in a secure manner. This association, while relying on short-term computational hardness assumptions, results in an increase of the practical security of QKD and eases its deployment.
\end{abstract}

\section{QKD, session authentication, and Digital Signatures}

Quantum Key Distribution (QKD) is a way to create shared and secret random values at both ends of a communication link, with a security guaranteed without computational hardness assumptions \cite{SCA09}. It requires however a classical authenticated channel, together with an untrusted 'quantum' channel (usually realized with an optical fiber or a free space optical transmission). This authenticated channel can be realized on top of an un-authenticated network connection using cryptographic primitives. The natural choice for these primitives is to use symmetric, unconditionally secure Message Authentication Codes like Wegman-Carter \cite{WC81}, Evaluation Hash \cite{MV84} or LFSR-based Toeplitz \cite{K94}.  Being symmetric, these primitives require a common secret; this is not a problem as soon as enough secret is created by the QKD link, but it is an undesirable constraint for the first run, as it forces the user to dispatch securely a common secret at both ends of the link. A very common argument against QKD is that, instead of exchanging a short common secret and using QKD to amplify it, one may as well exchange initially a very large secret and use it in place of the QKD output; the latter solution is easily realized thanks to the very low current price of storage. While this argument is not entirely correct,\footnote{Indeed, QKD is \emph{forward secure}, which means that each key produced is completely independent of past values; as a consequence, even an attacker having at some point in time a complete knowledge of the equipment state including its secrets, does not learn anything about future keys in a passive attack scenario. Contrary to the hard disk scenario where a one-time compromise is enough to obtain all the keys, QKD forces the attacker to perform a persistent active attack in order to obtain new keys, with a much higher risk of being detected.} it is desirable to have alternatives to the pre-sharing of a common secret.

Another argument against methods based on a common secret is that they are very hard to operate securely in practice. Indeed, the right way to implement them would be to store the secret on a device providing hardware security like a smart card (acting as a safe for the secret), but for this to be of interest the whole authentication tag computation needs to be performed inside the secure device. Unfortunately, the complete computation by a smart card of an authentication tag for a large set of messages corresponding to a QKD protocol run, typically consisting of several megabits, is unpractical. The secret must therefore be allowed to go out of the secure device; but then the very purpose of the secure device is defeated\footnote{One could think of a 2-stage scheme where the secure device authenticates a small digest of the message, but this can only be made secure in the computational sense: it must not be possible to find collisions for the function that transforms the message into the digest, and such collisions exist since the digest is smaller than the message.}.

Mitigation measures include enabling the secret to go out of the secure device only in a trusted environment, with a mechanism to authenticate the latter to the secure device like a pin code, or splitting the secret into several parts handled by independent parties, but the overall security assurance provided by these techniques does not compare favorably to the one offered by the resistance of cryptographic primitives, even computationally secure.

\medskip
Asymmetric cryptographic primitives which are used to negotiate keys in classical cryptography protocols usually employ computationally secure authentication means that are themselves asymmetric, that is, \emph{digital signature algorithms}. A common combination (normalized as ISO 9798-3 \cite{ISO9798}) is to use the Signed Diffie-Hellman algorithm, where a Diffie-Hellman key exchange is authenticated with digital signatures. The QKD protocol and the Diffie-Hellman algorithm are very similar in function, in that they both enable to create common secret values if a way to authenticate messages is available, although the security guarantees they provide differ. Digital signatures require the communicating parties to exchange \emph{public keys} in an \emph{authentic} way, contrary to symmetric MAC algorithms which require a common \emph{secret} value. This is a huge improvement because it is much easier to ensure that a value is authentic than it is to guarantee its secrecy\footnote{When there are more than two users, the separation of the key in a public and a private part also reduces the number of keys to distribute, since the same private-public key pair can be used to authenticate to several parties, the public part being distributed to all of them. In the point-to-point setting of QKD however, we are not concerned by this property.}. This is because a message alteration can be uncovered anytime \emph{after} it occurred, whereas preventing a loss of secrecy requires the perfect continuity of the protecting measures. Together with the invention of Public Key Infrastructures, this is what sparkled the success of public-key cryptography.

Similarly to the case of Diffie-Hellman, it is appealing to use an asymmetric signature algorithm to authenticate the first run of a pair of QKD equipments. Of course, doing this makes the QKD security depend on the strength of the signing algorithm, which reintroduces the very computational hardness assumptions QKD is supposed to be free of. 

In realistic deployments however, QKD will not be used stand-alone, encrypting traffic using one-time pad, and ensuring its integrity using unconditionally secure Message Authentication Codes, but rather together with computationally secure symmetric encryption and authentication algorithms built on top of symmetric ciphers like the AES \cite{AES}; in that setting, hardness assumptions are required to ensure the confidentiality and integrity of the user traffic, and therefore it makes sense to investigate the relationship between these assumptions and the ones backing the security of asymmetric signature algorithms.

\medskip
The existence of practical secure symmetric cryptography (stream ciphers, block ciphers, and hash functions) is equivalent to the existence of \emph{one-way functions}, that is, functions easy to compute and hard to invert. Indeed, block ciphers are pseudorandom permutations, i.e.\ permutations indexed by a key which are computationally indistinguishable from random permutations when the key is secret; stream ciphers are pseudorandom number generators; and hash functions are collision, $1^{\text{st}}$- and $2^{\text{nd}}$-preimage resistant functions\footnote{a function is $n^{\text{th}}$-preimage resistant if it is difficult to compute a $n^{\text{th}}$ preimage of a value $x$ given $n-1$ different preimages of $x$.}. One-way functions existence is known to be equivalent to the existence of pseudo-random number generators \cite{HILL93}, to the existence of pseudorandom functions \cite{GGM86} and to the existence of pseudorandom permutations \cite{LR88}. One-wayness is exactly $1^{\text{st}}$-preimage resistance; it is implied by $2^{\text{nd}}$-preimage resistance which is in turn implied by collision resistance; see \cite{DBLP:conf/fse/RogawayS04}. Conversely, a collision resistant, length-reducing function can be constructed from a one-way function \cite{R90}. Hence $2^{\text{nd}}$-preimage resistant or collision resistant functions exist iff one-way functions exist.

It is expected that one-way functions \emph{do} exist, although this conjecture, implying the famous conjecture $P \neq NP$, is not likely to be proven in the near future\footnote{for a general presentation of these issues, see chapter 9 of \cite{MOV}, and in particular remark 9.12}. The existence of one-way functions with a trapdoor, which are required to instantiate most asymmetric primitives, implies the existence of one-way functions, but there is no known converse result.
Assuming the existence of quantum computers, the security of the most well-known number-theory-based constructions used in asymmetric cryptography (RSA, discrete logarithm-based) collapse. On the contrary, one-way functions do not seem particularly threatened. The Grover quantum algorithm enabling exhaustive search with square root complexity \cite{G96} requires a key space of size $2^{2n}$ to ensure an exhaustive key search in $\gdo{2^n}$, as opposed to $2^n$ with classical computers, but this linear key length increase is manageable.

Of course, the situation with practical algorithms is more complex, since even if one-way functions and secure, logarithmic key size symmetric cryptography exist, it is not known whether the symmetric primitives used today are good approximations of their idealized counterparts. Symmetric ciphers and cryptographic hash functions like the SHA family \cite{SHA12rev} do not seem to rely on a small family of well-identified hypotheses of hardness of simple mathematical problems, unlike asymmetric algorithms\footnote{For instance, the RSA hypothesis - related to, and not stronger than factoring - for RSA \cite{RSA78}, the discrete logarithm in finite fields or elliptic curves for DSA/ECDSA \cite{DSArev3,JM99} and Schnorr Signatures \cite{Sch89}, or related problems like the Computational Diffie-Hellman problem, etc.}. This lack of structure has two consequences: there is no provable security reduction between symmetric algorithms, but conversely their security is not likely to collapse because of some sudden theoretical advance. The last 30 years of cryptanalytic progress showed that the security of symmetric primitives of early designs like DES \cite{DES} or hashing functions like SHA1 tend to erode slowly rather than abruptly, and that more mature designs (the AES competition contenders, the SHA2 family) exhibit a very good resistance to cryptanalysis. 

\medskip
A family of signature algorithms, Lamport signatures \cite{L79} and their derivatives, only require a function $f$ with $1^{\text{st}}$-preimage resistance (i.e.\ a one-way function) and a function with collision resistance $g$ to work.
\footnote{Since a length-reducing, collision resistant function can be built from a one-way function, such schemes show that the existence of one-way functions implies the existence of digital signatures \cite{R90}}.

The combination of QKD with public channel authentication performed with a secure instantiation of Lamport signatures is expected to have the same security properties as QKD combined with unconditional authentication methods. 
We propose to prove rigorously the security of this combination in the Universal Composability (UC) framework described in \cite{DBLP:conf/focs/Canetti01}, with the definitions of \cite{DBLP:journals/iacr/Canetti03} concerning the UC security of signatures. Indeed, Universal Composability is the right framework to securely assemble cryptographic primitives, either classical or quantum \cite{unruh04simulatable}. To sum up the previous discussion, the association of one-way-function-based signatures, and Lamport signatures in particular, with QKD  is desirable in practice compared to the use of one-way-trapdoor-based signature schemes because:
\begin{itemize}
\item
there is greater confidence in the security of one-way functions and collision-resistant functions used today than in one-way trapdoor constructions;
\item
the computational hardness assumptions that are needed regarding one-way functions and collision-resistant functions are  close to the ones needed to ensure the security of the mechanisms using the keys produced by QKD in most realistic scenarios; this is not the case for one-way-trapdoor security.
\end{itemize}

Whatever the computationally secure primitive used to ensure authentication, the hardness assumptions related to that primitive must only hold during the run time of a QKD session for the QKD protocol to remain forward-secure: no long-term hardness hypothesis is needed.

Finally this association is easier to deploy than QKD paired with an unconditional authentication scheme because the latter requires to deploy a common secret, a procedure which is difficult to implement properly.

\medskip
In the rest of this paper, we investigate in details this solution, examine some of its variants and detail the security properties obtained.

\medskip
\emph{Related work} : a more general survey of the relationship between various existence assumptions of classical cryptographic primitives including QKD is laid out in \cite{DBLP:conf/pqcrypto/IoannouM11}. Classical authenticated key exchange security models adapted to the QKD case are studied in \cite{cryptoeprint:2012:361}. 

Our contribution is to rigorously prove in the UC framework the security of the association of computationally secure signatures, and Lamport signatures in particular, with QKD.

\section{Lamport Signatures}

\subsection{Description}

For this paragraph, the main reference is the chapter 3 of the book \cite{BBD08}\footnote{available on-line at  \\ \url{http://www.cdc.informatik.tu-darmstadt.de/~dahmen/papers/hashbasedcrypto.pdf}}. 

\label{subsec:lamport_def}

$(n,\ell)$ Lamport signatures are $n \times \ell$-bit strings, where $n$ and $\ell$ are chosen according to the security requirements as explained in section \ref{subsec:dim}. 

As usual for digital signatures, a message to be signed is first transformed into a fixed-length string, its \emph{digest}, by a collision-resistant hash function $g$ chosen randomly in a collision-resistant function family $\mathcal{G}=\{g_k:\{0,1\}^* \to \{0,1\}^{\ell} |k\in K'\}$. The rest of the algorithm uses $f$ chosen randomly in a one-way function family $\{f_k:\{0,1\}^n \to \{0,1\}^n|k\in K\}$. 

In practice, $f$ and $g$ are often instatiated with a fixed cryptographic hash function, but the security analysis of collision-resistant and one-way function requires considering function families, for there exists trivial adversaries against the one-wayness or collision resistance of any fixed function.

\bigskip
\emph{Key generation algorithm\quad} Values $x_i[j] \in \{0,1\}^n,\ i=0,\ldots,\ell-1,\ j=0,1,$ are drawn randomly with the uniform distribution and make the private signature key. The public key is $\{y_i[j]=f(x_i[j]),\ i=0,\ldots,\ell-1,\ j=0,1\}$.

\bigskip
\emph{Signature algorithm\quad}
The signature of a message $M$ of digest $m=g(M)=m_0,\ldots, m_{\ell-1}$ is the $n \times \ell$-bit string $x_0[m_0],x_1[m_1],\ldots,x_{n-1}[m_{\ell-1}]$. 

\bigskip
\emph{Signature verification algorithm \quad}
The signature verification of a signature $s_0,\ldots, s_{\ell-1}$ of a message $M$ with digest $m=g(M)=m_0,\ldots, m_{\ell-1}$  outputs \texttt{True} if $\forall \ i=0,\ldots,\ell-1, \ f(s_i)=y_i[m_i] $, and \texttt{False} otherwise.

\subsection{The Case of QKD; one-time Signatures Usability}

The Lamport algorithm is not widely used because a key pair can only sign one message. Indeed, its security degrades very quickly when several messages are signed with the same key pair: this is to be expected since a signature is really just a part of the private key. More precisely, given $k$ signatures of messages whose digests are $m_i^j$, $i=0,\ldots,\ell-1$, $j=0,\ldots, k-1$, a signature for any message of digest $m'_0,\ldots,m'_{\ell-1}$ s.t. 
\begin{equation}
\label{eq:signable_msg}
\forall\, i\,,\ m'_i \in \{m_i^j|j=0,\ldots,k-1\}
\end{equation}
can be signed using the previous signatures. As soon as the signed messages digests differ on more that one bit (which occurs with overwhelming probability as soon as $k\geq 2$ since $\ell\gg 1$), new combinations of the differing bits can be used to create digests that can be signed with the publicly available data and that are different from the original message digests\footnote{for values of $k \ll n$, it may still be hard to find a message whose digest lies in the set of digests that can be signed using the revealed part of the private key, i.e.\ messages satisfying equation (\ref{eq:signable_msg}), but this property holds only for very small values of $k$ and cannot be used in practice.}. 

As we shall see in section \ref{merkle}, there are hash-based signature algorithms that are able to sign several messages with  a unique set of keys. However, in the QKD setting, the limitation to one signature is not an issue for two reasons:
\begin{itemize}
\item
The algorithm is only used to authenticate the first protocol run of a pair of QKD devices; subsequent executions are authenticated normally with a symmetric MAC algorithm using some of the common secret produced by the QKD link itself.
\item
To enable a recovery when this run was not successful, it is possible to include in the signed message a new public key that will be used to authenticate a new execution if needed. Additionally, a computational symmetric MAC can be used to check for message authenticity before using the Lamport mechanism: this will eliminate most failures before resorting to the signature and consuming the Lamport key, while not needing to put too much trust in the  symmetric key used by the computational MAC mechanism.
\end{itemize}

\subsection{UC Security of Lamport Signatures}

\label{subsec:secprop}

In this section we prove the UC security of the Lamport signature algorithm, from the more classical security notion of existential unforgeability under chosen-message attack (EU-CMA) \cite{Goldwasser:1988:DSS:45474.45480,DBLP:journals/iacr/Canetti03}, which models the ability for an adversary of a signature scheme to produce a forged signature for a message of her choice after interacting with the signature algorithm, and in particular after the observation of valid signatures of chosen messages. For the reader convenience, an informal definition of EU-CMA security is stated below.

\begin{defi}[EU-CMA security]
A EU-CMA adversary against a signature scheme (defined by its key generation, signature, and signature verification algorithms) is a randomized algorithm interacting 
\begin{itemize}
\item
with a key generation oracle which outputs identities-public key pairs,
\item
with a signature oracle which outputs valid signatures given on input a message to sign and one of the identities created with the key generation oracle, 
\item
with a corruption oracle which outputs the private key and the internal state of the signing algorithm when given on input one of the identities created with the key generation oracle.
\end{itemize}

A signature scheme is $(t,\epsilon)$ EU-CMA secure if no EU-CMA adversary can produce with a probability $\geq \epsilon$ and time $\leq t$ a valid message-signature pair $(M,\sigma)$ for a message $M$ never submitted to the signature oracle and for an uncorrupted identity.
\end{defi}

\begin{defi}[EU-CMA security, bounded queries]
A signature scheme is $(t,\epsilon,k)$ EU-CMA secure if $(t,\epsilon)$  EU-CMA secure against adversaries that are restricted to performing at most $k$ signature queries for each identity. It is said to be $(t,\epsilon,k,r)$ secure if it is $(t,\epsilon,k)$ EU-CMA secure against adversaries restricted to perform at most $r$ key generation queries.
\end{defi}

It was seen that $(t,\epsilon,k)$ EU-CMA security of Lamport signatures breaks down if $k > 1$. We prove conversely that it is $(t,\epsilon,1)$ EU-CMA secure when correctly instantiated.

\begin{defi}[one-way function family]
A family of functions $\mathcal{F}=\{f_k:\{0,1\}^n \to \{0,1\}^n |k\in K\}$ for some $n \in \N$, some finite set $K$ is $(t,\eps)$ one-way if the probability for any randomized algorithm given on input a random $k\in K$ and $y \in f_k(\{0,1\}^n)$ to output in time $\leq t$ $x$ s.t. $f_k(x)=y$ is below $\eps$.
\end{defi}

\begin{defi}[collision-resistant function family]
A family of functions $\mathcal{G}=\{g_k:\{0,1\}^* \to \{0,1\}^{\ell} |k\in K\}$ for some $\ell \in \N$, some finite set $K$ is $(t_{\text{CR}},\epsilon_{\text{CR}})$ collision-resistant if the probability for any randomized algorithm given on input a random $k\in K$  to output in time $\leq t_{\text{CR}}$ $x,x'$ s.t. $g_k(x)=g_k(x')$ is below $\epsilon_{\text{CR}}$.
\end{defi}

The EU-CMA security proof below is similar for instance to the proof found in \cite{coursB} of a simplified Lamport signature scheme. There,  a slightly weaker EF-CMA security model is used, where the attacker must produce a valid signature for some fixed public key.

\begin{theo}[EU-CMA security of the full Lamport signature scheme]
Let $n,\ell \in \mathbb{N}$, $\mathcal{F}=\{f_k:\{0,1\}^n \to \{0,1\}^n |k\in K\}$ be a family of $(t,\epsilon_{\text{OW}})$ one-way functions, and $\mathcal{G}=\{g_k:\{0,1\}^* \to \{0,1\}^{\ell} |k\in K'\}$ be a family of $(t,\epsilon_{\text{CR}})$ collision-resistant functions. Then the $(n,\ell)$ Lamport signature scheme $L(\mathcal{F},\mathcal{G})$ that uses $\mathcal{F}$ and $\mathcal{G}$ is $(t,\eps_{\text{L}}, 1)$ EU-CMA secure with the additional restriction on attackers that they are allowed at most $r$ queries to the key generation oracle, for $\eps_{\text{L}} = r(2\, \ell\, \epsilon_{\text{OW}} + \eps_{\text{CR}})$.
\label{theo:lamp-eu-cma}
\end{theo}

\proof

The proof proceeds by transforming an attacker $\mathcal{A}$ against the signature scheme into an attacker against the underlying one-way function family or the collision-resistant function family. Given a challenge $(k,y)$ for the one-way function family $\mathcal{F}$  and a challenge $k'$ against the collision-resistant function family $\mathcal{G}$, the environment of the attacker is simulated as follows. $(k,k')$ is given to the attacker. Some indexes $r', \ell', b$ with index $0\leq r' < r$, $0\leq \ell' < \ell$ and $b\in\{0,1\}$ are chosen randomly with the uniform distribution. As will be seen later $r'$ is a guess on the identity attacked by $\mathcal{A}$ . The key generation and signature oracles are simulated as in the Lamport definition with $f=f_k$ and $g=g_{k'}$, except for the $r'$-th key generation query (if it is made by the attacker) where $y_{\ell'}[b]$ is set to $y$. No corresponding private key value $x_{\ell'}[b]$ is therefore known by the simulator. 

If a signature query is made on identity $r'$, the simulator is able to answer it with probably 1/2; it aborts the simulation otherwise. If a corrupt query is issued on identity $r'$, the simulation is also aborted. All other queries succeed with probability 1.

Assume the attacker succeeds and outputs a valid forgery $(M_1,\sigma)$ for identity $r''$. If $r''\neq r'$, the simulation is aborted. Otherwise, three cases can occur.
\begin{itemize}
\item
There was no signature query made on identity $r'$. In that case, with probability $1/2$ $g(M_1)_{\ell'} = b$ and $\sigma$ contains a preimage of $y$. 
\item
A signature query was made for some message $M_0 \neq M_1$ and $g(M_0) \neq g(M_1)$. In that case, with probability $1/\ell$, $g(M_1)_{\ell'} =b=1-g(M_0)_{\ell'} $ and $\sigma$ contains a preimage of $y$.
\item
A signature query was made for some message $M_0 \neq M_1$ and $g(M_0) = g(M_1)$. Then $(M_0,M_1)$ is a collision pair for $g$.
\end{itemize}

In any of these three cases, if the simulator gets a collision on $g$ or a preimage of $y$ by $f$ it outputs it and stops the simulation. Let $\eps_{\on{pre}'}$ the probability of the first two cases and $\eps_{\on{coll}}$ the probability of the last case. If $r'$ is the correct guess on the identity attacked by $\mathcal{A}$, which is the case with probability $1/r$, no corrupt query happens on identity $r'$ before the forgery is produced. Thus if $\eps$ is the success probability of $\mathcal{A}$,
\[\eps_{\on{pre}'} + \eps_{\on{coll}} = \eps / r.\]
The success probability $\eps_{\on{pre}}$ of the simulator against $\mathcal{F}$  satisfies $\eps_{\on{pre}} \geq \eps_{\on{pre}'}/2\ell$. The success probability of the simulator against $\mathcal{G}$ is  $\eps_{\on{coll}}$. Finally if the $\mathcal{A}$ runs in time $t$, the simulator also runs in time $t$ and

\[\eps \leq r(2\ell \eps_{\on{pre}} +\eps_{\on{coll}}) \leq r(2\ell \eps_{\on{OW}} +\eps_{\on{CR}})\]
\endproof

To prove the UC security of Lamport signatures, we use a non-asymptotic version of theorem 2 of \cite{DBLP:journals/iacr/Canetti03}, which ensures that it is equivalent to realise the ideal functionality of signature $\mathcal{F}_{\on{SIG}}$ (the UC definition of signature security), and to be EU-CMA secure. Both the EU-CMA notion and the corresponding UC ideal function can be restricted to the $k$-signature setting, where an adversary is only allowed to invoke the signing oracle $k$ times per identity. Let us call the UC restricted functionality $\mathcal{F}_{\on{SIG}k}$.

\begin{theo}[equivalence between EU-CMA security and UC security of signatures $\mathcal{F}_{\on{SIG}}$]
Assume that a signature scheme $\pi$ is $(t, \eps)$ (resp. $(t, \eps, k)$, $(t, \eps, k, 1)$) EU-CMA secure. Then for any adversary $\mathcal{A}$ running in time $t$ and performing at most $r$ key generation queries, there exists a simulator $\mathcal{S}$ such that the advantage $\eps_{\on{UC}}$ of any environment $\mathcal{Z}$ to distinguish whether it is interacting with $\mathcal{F}_{\on{SIG}}$ (resp. $\mathcal{F}_{\on{SIG}k}$, $\mathcal{F}_{\on{SIG}k}$) and $\mathcal{S}$ or with $\pi$ and $\mathcal{A}$ satisfies $\eps_{\on{UC}} \leq r \, \eps$.

We say that $\pi$ is a $(t,r \eps,r)$ realization of functionality  $\mathcal{F}_{\on{SIG}}$ (resp. $\mathcal{F}_{\on{SIG}k}$, $\mathcal{F}_{\on{SIG}k}$).
\label{theo:eq-eu-cma-f-sig1}
\end{theo}

\proof 
As theorem 2 of \cite{DBLP:journals/iacr/Canetti03}. The argument for the bound on the distinguishing probability comes from two observations: 
\begin{itemize}
\item
To enable the environment to distinguish the real and the ideal case with nonzero advantage, the adversary must produce a valid forged signature during its execution, enabling to construct an attacker $G$ in the EU-CMA model from a pair $(\mathcal{A}, \mathcal{Z})$ distinguishing the real and the ideal case;
\item
The attacker $G$ above needs to guess in advance the identity corresponding to the forgery produced. This guess succeeds with probability $1/r$. This induces a loss factor $r$ on the reduction.
\end{itemize}

The number of signature queries made by $\mathcal{A}$ in the $\mathcal{F}_{\on{SIG}}$ setting and by the forger $G$ in the EU-CMA setting are the same which ensures that $(\cdot,\cdot,k)$ EU-CMA secure schemes and realizations of $\mathcal{F}_{\on{SIG}k}$ functionality are in correspondence in this proof.

Furthermore, the forger $G$ simulates internally all identities except the one it attempts to forge a signature for. Therefore $G$ performs only one key generation query on $\pi$ in the EU-CMA model no matter $r$.

\endproof

We can now derive our main result regarding Lamport UC security.

\begin{coro}[UC security of the Lamport signature scheme]
\label{coro:lamport-uc}
Let $n,\ell \in \mathbb{N}$, $\mathcal{F}=\{f_k:\{0,1\}^n \to \{0,1\}^n |k\in K\}$ be a family of $(t,\epsilon_{\text{OW}})$ one-way functions, and $\mathcal{G}=\{g_k:\{0,1\}^* \to \{0,1\}^{\ell} |k\in K\}$ be a family of $(t,\epsilon_{\text{CR}})$ collision-resistant functions.

Then the $(n,\ell)$ Lamport scheme $L(\mathcal{F},\mathcal{G})$ is a 
\[(t,r(2\, \ell\, \epsilon_{\text{OW}} + \eps_{\text{CR}}),r)\]
realization of functionality  $\mathcal{F}_{\on{SIG1}}$.
\end{coro}

\proof
Direct composition of theorems \ref{theo:lamp-eu-cma} with $r=1$ and \ref{theo:eq-eu-cma-f-sig1}.
\endproof

In the UC setting, the signature functionality is a superset of the message authentication functionality: the ideal functionalities differ only by the possibility for the adversary to query public keys which exists only for signatures. Therefore a signing scheme can securely be composed with a composable protocol expecting a message authentication scheme, such as a QKD scheme having universal security as defined in \cite{Ben-or05theuniversal}. This is formalized in the result below.

\begin{theo}[Composition result for QKD and $\mathcal{F}_{\on{SIG1}}$ signatures]
Assume that a QKD protocol $A$ paired with classical authentic channels satisfies the composable security property
\[\frac{1}{2} \left \| \rho_{SE}^{A} - \rho_U \otimes \rho_{E}^{A} \right \|_1 < \eps \]
where $\rho_{SE}$ represents the joint quantum state of an attacker and the final secret key, $\rho_U$ is the fully mixed state in the Hilbert space corresponding to the key space, and $\rho_E$ the attacker state. Then:
\begin{enumerate}
\item
Assume protocol parties have access to an unauthenticated communication channel in addition to an authenticated channel. Then the protocol can be transformed into a protocol where the authenticated channel is used to transmit only one message in each direction, with no security loss, i.e. one still has for the transformed protocol
\[\frac{1}{2} \left \| \rho_{SE}^{A} - \rho_U \otimes \rho_{E}^{A} \right \|_1 < \eps \]
\item
If this transformed protocol is composed with a $(t,\eps',2)$ realization of functionality $\mathcal{F}_{\on{SIG1}}$ to authenticate the two exchanged messages, where any adversary running in time $\leq t$ can distinguish between the ideal functionality and the realization with probability $\leq \eps'$, the resulting protocol $B$ is $\eps + \eps'$-secure against adversaries running in time $\leq t$:
\[\frac{1}{2} \left \| \rho_{SE}^{B} - \rho_U \otimes \rho_{E}^{B} \right\|_1 < \eps + \eps'\]
\item
When the $\mathcal{F}_{\on{SIG1}}$ functionality is provided by the $(n,\ell)$ Lamport algorithm instantiated with a $(t,\epsilon_{\text{OW}})$ one-way function family and a $(t,\epsilon_{\text{CR}})$ collision-resistant function family,
\[\frac{1}{2} \left \| \rho_{SE}^{B} - \rho_U \otimes \rho_{E}^{B} \right\|_1 < \eps + 4\, \ell\, \epsilon_{\text{OW}} + 2\,\eps_{\text{CR}}\]
\end{enumerate}
\end{theo}

\proof
\begin{enumerate}
\item
The original protocol is transformed as follows. Each message is sent on the unauthenticated channel. When some of the parties of the protocol has finished  its execution and is ready to produce some key, it first builds an unambiguous description of all the messages exchanged during the protocol run (the "conversation"). The exact nature of this description depends on the properties of the authenticated channel used by the unmodified protocol: for instance, if it guarantees an ordering of the messages sent in both directions, the description of the conversation should include that ordering information. This description is then sent on the authenticated channel. Conversely, each party  waits for the reception of this description from its peer and checks that it matches its own view of the conversation. If not, the protocol is aborted. If no party aborts, then each message sent was received unmodified, no message was deleted and the message ordering properties guaranteed by the authenticated channel model were preserved. Therefore for each successful execution of the protocol, everything happens as if an authenticated channel was used and the security property of the unmodified protocol holds.
\item
The protocol modified to use the authenticated channel only once in each direction is composed with functionality $\mathcal{F}_{\on{SIG1}}$ to perform the message authentications. Each identity is only used once therefore the usage of functionality  $\mathcal{F}_{\on{SIG1}}$ is correct. The result is then a consequence of the composition theorem for UC protocols.
\item
Apply corollary \ref{coro:lamport-uc} with $r=2$. 
\end{enumerate}
\endproof

\subsection{Parameter Dimensioning}

\label{subsec:dim}

Generic (non quantum) attacks against a hash function producing $\ell$-bit hashes enable to find collisions in $\gdo{2^{\ell/2}}$ hash function computations. For a $n$-bit hash function, preimages are found in a generic way in $\gdo{2^n}$ hash function computations. $k$-bit classical security (i.e.\ best non-quantum attack in $\gdo{2^k}$ operations) for Lamport signatures therefore requires $n\geq k$ and $\ell\geq 2k$. Typically we want 128-bit security, which yields $n \geq 128$ bits and $\ell \geq 256$ bits.

If quantum generic attacks are taken into account, the picture changes a bit. Finding preimages of an $n$-bit hash function can be performed in $\gdo{2^{n/2}}$ operations using Grover algorithm, and there is a quantum algorithm with complexity $\gdo{2^{\ell/3}}$ able to find with good probability a collision of $\ell$-bit hashes \cite{BHT97}. However it requires  a (quantum) memory of size $\gdo{2^{\ell/3}}$ \cite{gr04}, so that we rather include its analysis to the next paragraph about parallel methods. With a constant or log amount of memory, the best known quantum attacks for $n$-bit hash preimage and $\ell$-bit hash collision have complexity $\gdo{2^{n/2}}$ and $\gdo{2^{\ell/2}}$, respectively. For $k$-bit security, one should therefore choose $\ell,n\geq 2k$.

Assuming some parallelism, results are again different. With a (classical or quantum, computation or memory) resource size of $\gdo{2^\mu}$, and realistic communication models, the best known generic preimage complexity for a $n$-bit hash function is $\gdo{2^{(n-\mu)/2}}$ and the best collision attack for a $\ell$-bit hash function, $\gdo{2^{\ell/2-\mu}}$  \cite{B09}. Hence one should choose  $n \geq 2k + \mu$ and $\ell \geq 2 (k+\mu)$. Assuming $k=128$ and $\mu=64$ (which is an extremely large security margin), this gives $n \geq 320$ bits  and $\ell \geq 384$ bits.

The complexities above are given for an attacker with success probability one. Going back to the $\epsilon', \epsilon''$ above, the success probability of an attacker using only a fraction $\eta$ of the resources indicated has a success probability $\eta^2$, except for the classical preimage search algorithm where the probability scales linearly with the effort.

\subsection{Operation with a Secure Device}

A \emph{secure device} provides facilities to store data in a confidential and/or authentic way and to perform operations using this data. A typical cheap secure device is a smart card. It has a limited computational power, but is designed to store  keys securely and enable to operate some cryptographic algorithms making use of these keys, typically a hashing algorithm (usually SHA1 or SHA2) and a block cipher algorithm (3DES or AES). More powerful secure devices ('Hardware Security Modules') can sit in computers but usually offer security assurances lower than smart cards.

The right way to implement a public-key algorithm such as Lamport is to never let private keys go out of secure devices. Therefore each secure device, at each end of the QKD link, generates its own private key and discloses the corresponding public key. During a trusted initialization phase, each device then receives the public key of the other device, or some digest of it, enabling it to later authenticate the signature of the other party. 

Both the collision-resistant function and the preimage-resistant function are implemented with the hash function provided by the secure device.

Since the first step in computing a signature (applying $g$) does not involve the private key, and results into a small-fixed-length string, this step can be performed outside the secure device. The hash is then provided to the secure device. With notations of section \ref{subsec:lamport_def}, the latter then picks the values $x_0[m_0],x_1[m_1],\ldots,x_{\ell-1}[m_{\ell-1}]$ corresponding to the digest $m_0,\ldots,m_{\ell}$ provided.

To ensure a correct usage of the algorithm as well as to effectively protect the private key, the sequencing of the instructions to the secure device should prevent multiple signing with the same private key. For instance, if the values $x_i[m_i]$ are output successively by the secure device, as it is the case with a smart card that cannot output the signature all-at-once, the private key should be erased as soon as the last signature part is output, or even better, $x_i[1-m_i]$ can be erased as soon as $m_i$ is known by the secure device.

\section{Variants}

\subsection{Reducing the Private and Public Key Size}

\label{subsection:prng}

An issue with the Lamport algorithm is its large key size: for instance, for $\ell =256$ and $n=128$, the public key size is $2 \times 256 \times 128  = 64$Kb. With the more conservative parameters of paragraph \ref{subsec:secprop}, the size becomes $2 \times 384 \times 320 = 240$Kb. Such a key cannot be stored on most smart cards.

A standard way to overcome this is to generate the private key from a pseudo-random number generator (PRNG) and to store only the secret $s$ of the PRNG; it is then possible to compute the public and the private key on-the-fly. The public key and signature are then computed and output piece by piece, typically one hash at a time.

Security wise, using this construction requires to take into account the PRNG security. 

\begin{defi}[PRNG security, adapted from \cite{GGM86}]
Let $k \leq \ell \in \N$. A $(k,\ell)$-PRNG is a function $h: \{0,1\}^k \to \{0,1\}^{\ell}$. 
A distinguishing adversary $\mathcal{A}$ against $h$ plays the following game. Random values $s \in \{0,1\}^k$, $T \in \{0,1\}^{\ell}$, and $b\in \{0,1\}$ are drawn uniformly. $\mathcal{A}$ is given $c=h(s)$ if $b=0$, and $c=T$ otherwise; it must guess $b$.
$h$ is a $(t_{\text{RG}},\epsilon_{\text{RG}})$-secure PRNG if no randomized distinguishing adversary running in time $\leq t_{\text{RG}}$ has advantage $\eps=|2 \times p - 1|$ larger than $\eps_{\text{RG}}$, where $p$ is its probability to guess $b$ correctly.
\end{defi}

The security loss incurred by the use of a PRNG to generate the Lamport private key is quantified in the result below.

\begin{theo}[EU-CMA security of the full Lamport signature scheme with a PRNG-generated private key]
Let $n,\ell \in \mathbb{N}$, $\mathcal{F}=\{f_k:\{0,1\}^n \to \{0,1\}^n |k\in K\}$ be a family of $(t,\epsilon_{\text{OW}})$ one-way functions, $\mathcal{G}=\{g_k:\{0,1\}^* \to \{0,1\}^{\ell} |k\in K'\}$ be a family of $(t,\epsilon_{\text{CR}})$ collision-resistant functions, and $h$ a $(k, 2\,n\,\ell)$ $(t_{\text{RG}},\epsilon_{\text{RG}})$-secure PRNG. Then the $(n,\ell)$ Lamport signature scheme $L(\mathcal{F},\mathcal{G})$ that uses $\mathcal{F}$,  $\mathcal{G}$ and $h$ for the private key generation is $(t,r(2\, \ell\, \epsilon_{\text{OW}} + \eps_{\text{CR}} + \eps_{\text{RG}}), 1, r)$ EU-CMA secure.
\label{theo:lamp-eu-cma-prng}
\end{theo}

\proof
Let $\mathcal{A}$ be an EU-CMA attacker against a $(n,\ell)$ Lamport scheme, and running in time $\leq t$. Let $\eps_j$ be its success probability ia a EU-CMA game where the first $j$ identities are created with the PRNG and random seeds, and the others with random private keys. $\eps_0$ is the success probability against the unmodified Lamport scheme and there exists $i$ s.t. $|\eps_{i+1} - \eps_{1}| \geq 1/r |\eps_r - \eps_1|$.

From $\mathcal{A}$, construct an adversary against the PRNG as follows. The private key of the $i+1$-th identity created in the EU-CMA game is created from the PRNG challenge that is either a PRNG output or a random string. For $j\leq i$, private keys are created with a PRNG. For $j>i+1$, they are random. Output $b'$ where $b'=0$ iff $\mathcal{A}$ wins the EU-CMA game. Then $\on{Pr}(b'=0|b=0)-\on{Pr}(b'=0|b=1)=\eps_{i+1}-\eps_i$. It can be verified that the advantage of our attacker against the PRNG is $|\on{Pr}(b'=0|b=0)-\on{Pr}(b'=0|b=1)|$. As a consequence,
\[\eps_{\text{RG}} \geq |\eps_{i+1}-\eps_i|\geq 1/r |\eps_r - \eps_1|.\]

This, combined with theorem \ref{theo:lamp-eu-cma}, gives the final result.
\endproof

\begin{coro}[UC Security of the Lamport Signature Scheme with Private Key generated by a PRNG]
With the same assumptions and notations as theorem \ref{theo:lamp-eu-cma-prng}, the Lamport scheme is for any $r\in \N$ a \[(t,r(2\, \ell\, \epsilon_{\text{OW}} + \eps_{\text{CR}} + \eps_{\text{RG}}),r)\] realization of functionality $\mathcal{F}_{\on{SIG1}}$.
\end{coro}

\proof
Direct composition of theorems \ref{theo:lamp-eu-cma-prng} with $r=1$ and \ref{theo:eq-eu-cma-f-sig1}.
\endproof

The generic brute-force attack against a PRNG enumerates all seeds $s$ and outputs $b=0$ if and only if a seed $s$ such that $h(s)=c$ exists. This algorithm runs in time $\gdo{2^k}$ and wins with advantage $\geq 1-2^{k-\ell}$. This is similar to the preimage attacks we discussed in paragraph \ref{subsec:dim}, and the bounds given there for $n$ apply for the size $k$ of $s$. Secure PRNG constructions from secure hashing primitives and satisfying these size constraints exist (see for instance the annex of \cite{DSArev3}).

\subsection{Enabling Multiple Signatures: Merkle trees}
\label{merkle}

Lamport signatures, or, for the matter, any one-time signature (OTS) scheme, can be paired with \emph{Merkle trees}, which are a construction using only preimage-resistant functions and enabling the authentication of a large family of public signature keys with only one short value. We omit a complete description of Merkle trees, which is detailed for instance in \cite{BBD08}, but only describe their role and associated cost.

A Merkle tree of depth $H$ uses a collision-resistant function $T: \{0,1\}^{*} \to \{0,1\}^m$, and enables to perform $2^H$ signatures. The tree 'public key' is a $m$-bit value.

Each message signature consists in a public key and a signature by the underlying one-time signature scheme, with some added information value enabling to authenticate the OTS public key. This added information consists in $H$ $m$-bit values. The signature verification, aside from the underlying OTS signature verification, requires $H$ computations of $m$.

The signature algorithm calls the underlying OTS algorithm once and requires additionally $\gdo{H}$ computations of $T$ and the storage of $\gdo{H}$ values of $T$. 

Initially, $2^H$ OTS public-private key pairs must be generated. This is usually too large to store; instead, a PRNG can be used, with ideas similar to those of paragraph \ref{subsection:prng}. The computation of the tree public key requires $2^H - 1$ computations of $T$ and uses the $2^H$ underlying OTS public keys, which are themselves computed in the case of the Lamport algorithm through $2^{H+\ell+1}$ calls to $f$.

The success probability of an attacker against this scheme in the EF-CMA model, attempting to forge a signature after requesting at most $2^H$ signatures, is bounded by $2^H \eps_{OTS} + \eps_{CR}$ where $\eps_{OTS}$ is the success probability of an attacker against the underlying OTS scheme, and $\eps_{CR}$ the success probability of an attacker against the collision resistance of $T$. The output size $t$ of $T$ should therefore be chosen as $\ell$ for the Lamport signature scheme. For a proof of this result and more discussion on the attackers running times, see theorem 8.2 of \cite{BBD08}.

In the UC sense, there is a similar result:
\begin{theo}[UC security of one-time signatures combined with Merkle trees construct]
The combination of a $(t,\eps_{OTS},1)$ realization of functionality $\mathcal{F}_{\on{SIG1}}$, and of the Merkle tree construction of depth $H$ using a $(t,\eps_{CR'})$ collision-resistant function family is for any $r,H\in \N$ a $(t, r(2^H \, \eps_{OTS} + \eps_{CR'}),r)$ realization of functionality $\mathcal{F}_{\on{SIG}2^H}$.
\end{theo}

\proof
EU-CMA security is proved as in theorem \ref{theo:lamp-eu-cma}: one first guesses the identity for which the forgery will occur, then a challenge for the underlying OTS scheme is injected in one of the $2^H$ OTS public keys of this identity; if the guess is right, a correct answer of an EU-CMA attacker against the composition of the OTS scheme and Merkle tree construct produces a correct answer for the underlying OTS challenge with probability $2^{-H}$, or a collision on $T$. Then theorem \ref{theo:eq-eu-cma-f-sig1} is applied with $r=1$.
\label{theo:ots-merkle}
\endproof

\begin{theo}[UC security of Lamport signatures Combined with Merkle Trees Construct]
The combination of the Lamport scheme with the hypotheses of theorem \ref{theo:lamp-eu-cma} and of Merkle trees of depth  $H$ built with a $(t,\eps_{CR'})$ collision-resistant function family is a $(t,r\left[2^H(2\, \ell\, \epsilon_{\text{OW}} + \eps_{\text{CR}}) + \eps_{\text{CR'}}\right],r)$ realization of functionality $\mathcal{F}_{\on{SIG}2^H}$.
\end{theo}

\proof
Composition of theorems \ref{theo:lamp-eu-cma} with $r=1$ and \ref{theo:ots-merkle}.
\endproof

For the sake of completeness, let us formulate a result for the combination of Lamport Signatures and Merkles trees, when the $2^H$ Lamport private keys for each identity of the Lamport scheme are generated from one single seed using a PRNG.

\begin{theo} [UC security of Lamport signatures combined with Merkle trees and PRNG]

Let $m,n,\ell, H \in \mathbb{N}$ and
\begin{itemize}
\item
$\mathcal{F}=\{f_k:\{0,1\}^n \to \{0,1\}^n |k\in K\}$ be a $(t,\epsilon_{\text{OW}})$ one-way function family,
\item
$\mathcal{G}=\{g_k:\{0,1\}^* \to \{0,1\}^{\ell} |k\in K'\}$ be a $(t,\epsilon_{\text{CR}})$ collision-resistant function family,
\item
$\mathcal{T}=\{T_k:\{0,1\}^* \to \{0,1\}^{m} |k\in K''\}$ be a $(t,\epsilon_{\text{CR'}})$  collision-resistant function family,
\item
$h$ a $(k, 2^H(2\,n\,\ell))$ $(t_{\text{RG}},\epsilon_{\text{RG}})$-secure PRNG.
\end{itemize}
Then the combination of the $(n,\ell)$ Lamport signature scheme that uses $\mathcal{F}$,  $\mathcal{G}$ with depth $H$ Merkle trees, where each private key of the combination is generated with $h$ from a random seed, is a 
\[(t,r\left[2^H(2\, \ell\, \epsilon_{\text{OW}} + \epsilon_{\text{RG}} + \eps_{\text{CR}}) + \eps_{\text{CR'}}\right],r)\] realization of functionality $\mathcal{F}_{\on{SIG}2^H}$.
\end{theo}

\proof
As the proof of theorem \ref{theo:lamp-eu-cma-prng}, starting from the EU-CMA security of the combination of the Lamport algorithm and Merkle trees.
\endproof

In the usage context of quantum key distribution, the typical instantiation of Merkle trees would use a rather small parameter $H$ (say $H<10$) to avoid sending a new public key in each signature, while retaining a small signature and verification overhead (this is especially true on the verification side where the Lamport verification alone requires $\ell$ hash computations). This also ensures the provable security loss factor incurred, $2^H$, is small. 


\section{Conclusion}

We have seen that provided some minimal hypotheses are fulfilled, namely the existence of preimage-resistant and collision-resistant functions, some signatures schemes well known in classical cryptography can be securely composed with a QKD protocol to authenticate communications on a classical channel without a common secret, and thereby enable to bootstrap key production on a quantum key distribution link. In practical scenarios, collision-resistant and preimage-resistant functions are instantiated with a cryptographic hash function. We contend that the security gained by the removal of the initial common secret far outweights the loss caused by the dependence to the preimage and collision resistance of the hash function used, notably because these properties are required to hold only until the initial QKD protocol run finishes, and not as long as the keys produced by the QKD link are supposed to remain secret.

We believe that there are other ways to use properties related to one-wayness, like pseudo-randomness, to improve the practicality and/or security of QKD protocols, without sacrificing its most fundamental property, the forward secrecy.

\bibliography{biblio}
\iftoggle{ieeeversion}
{%
\bibliographystyle{plain}
}{%
\bibliographystyle{alpha}
}

\end{document}